\documentclass[11pt]{article}
\usepackage{moriond,epsfig}

\def\be{\begin{equation}}
\def\ee{\end{equation}}
\def\bea{\begin{eqnarray}}
\def\eea{\end{eqnarray}}

\begin{document}
\vspace*{4cm}
\title{Long distance modifications of gravity in four dimensions.}

\author{I. NAVARRO$^*$ and K. VAN ACOLEYEN$^{\dagger}$}

\address{$^*$DAMTP, University of Cambridge, CB3 0WA Cambridge, UK\\
$^{\dagger}$ IPPP, University of Durham, DH1 3LR Durham, UK }

\maketitle\abstracts{We discuss some general characteristics of
modifications of the 4D Einstein-Hilbert action that become
important for low space-time curvatures. In particular we focus on
the chameleon-like behaviour of the massive gravitational degrees of
freedom. Generically there is at least one extra scalar that is
light on cosmic scales, but for certain models it becomes heavy
close to any mass source.}

\section{The models}
In this talk we will look at some aspects of modifications of the
four dimensional Einstein-Hilbert action that are of the following
general form:
\bea S=\int \!\!d^4x\sqrt{-g}\frac{1}{16\pi
  G_N}\left[R +F(R,P,Q)\right]\,,\label{action}\eea
with $P \equiv R_{\mu\nu}R^{\mu\nu}$ and $Q \equiv
R_{\mu\nu\lambda\rho}R^{\mu\nu\lambda\rho}$. We want to modify
gravity at long distances or low curvatures, so we introduce some
{\em crossover scale} $\mu$ where the modification kicks in, such
that we have conventional general relativity (GR) for large
curvatures and something different for low curvatures: \be R\gg
F(R,P,Q)\,\,\textrm{for} \,\,R^2\!,P,Q\gg
\mu^4\,\,\,\,\,\,\,\,\textrm{and}\,\,\,\,\,\,\,\,R \ll
F(R,P,Q)\,\,\textrm{for} \,\,R^2\!,P,Q \ll \mu^4\,.
\label{highcurvature}\ee Models of this kind have been proposed as a
way to generate late time acceleration without a cosmological
constant\cite{Capozziello:2003tk}. For this to work the crossover
scale has to be of the order of today's Hubble constant: $\mu \sim
H_0$. In the Friedmann equation obtained for these theories, the
modification then only becomes important at the present stage of the
expansion of the Universe. Obviously the detailed predictions will
depend on the specific model at hand. So far the only model that has
been put to test is the $n=1$ case of: \be
F(R,P,Q)=-\frac{\mu^{2+4n}}{(aR^2+b P + c
Q)^n}\,,\label{specmodels}\ee and it was found that it can fit the
SN data\cite{Mena:2005ta} for a certain range of parameters $a,b,c$
(but see\cite{DeFelice:2006pg} for stability constraints).

\section{The excitations}
To understand the physics of the models (\ref{action}) a good
starting point is to examine their excitations. In the case of GR,
the two only degrees of freedom are contained in the massless spin 2
graviton. For a massless spin 2 particle the weak field limit is
unique and this mediates the gravitational force in a very specific
way. For the models (\ref{action}),  one will have six more degrees
of freedom in addition to the massless spin 2
graviton\cite{Hindawi:1995cu}. On vacuum, one of them is a massive
scalar and the other five are contained in a massive spin 2 ghost,
whose negative kinetic energy would arguably lead to the decay of a
homogeneous background into a complete inhomogeneous state full of
negative and positive energy excitations. To check the validity of
these type of models one could try to calculate the decay time of
the vacuum and check if the result is compatible with observations,
but it is probably safer not to have ghosts at all.

Fortunately one can show that the massive spin 2 ghost disappears
altogether for modifications of the form $F(R,Q-4P)$
\cite{Navarro:2005da,Comelli:2005tn} , and we will assume this form
from now on. The modification is then characterized by one extra
scalar degree of freedom in addition to the massless spin 2 graviton
of GR. Remember that by the conditions (\ref{highcurvature}) we can
neglect the corrections due to the modification for large curvatures. This will translate
itself to a large mass of the extra scalar on backgrounds that have
a large curvature. Indeed, since the interaction range of an
excitation is typically of the order of the inverse mass (we use
units such that $c=\hbar=1$), the scalar will effectively
decouple on those backgrounds.  For the models (\ref{specmodels})
for instance, the running of the mass $m_s$ with the
background curvature $\mathcal{R}$ is given by an expression like
\cite{Navarro:2005da}:
\be m_s^2\sim \mathcal{R}\left(\frac{\mathcal{R}}{\mu^2}\right)^{2n+1}\,,
\label{runningmass} \ee where $\mathcal{R}$ stands for a certain
combination of components of the background Riemann tensor. Also the
effective Newton's constant, describing the coupling of the
gravitational excitations to matter, will run with the background
curvature: \be G^{eff}_N\sim
\frac{G_N}{1+(\frac{\mu^2}{\mathcal{R}})^{2n+1}}\,,\ee and again
we see that we recover GR for large curvatures
($G^{eff}_N\rightarrow G_N$).

This background dependence of the mass of the scalar and of the
effective Newton's constant is a manifestation of the violation of
the strong equivalence principle for this type of theories: the
properties of the local gravitational excitations depend
intrinsically on the background, in that sense they behave
like a gravitational chameleon\cite{Khoury:2003aq}. In general, one
can have a complete breakdown of the local Lorentz symmetry for the
short distance excitations. To assess the stability of a
certain background under short distance fluctuations, one will have
to look at the propagation of the degrees of freedom {\em on
that background}. On general FRW backgrounds for instance, the
propagation of the spin 2 graviton can be sub- or superluminal and
its kinetic energy can be positive or negative\cite{DeFelice:2006pg}.

\section{The Schwarzschild solution}
Gravity, like electromagnetism, has infinite range. This means that GR influences a huge variety of
phenomena on a vast range of distance scales. So if you fiddle with
it to get some interesting modification for the expansion of the
Universe, you should check the effect on all the other gravitational
phenomena. Now, if we think about the modification in terms of extra
degrees of freedom there seems to be a problem. On one hand, if
we want a modification at cosmic scales, we need the mass of these
extra degrees of freedom to be at most of the order of today's
Hubble constant, giving them an effectively infinite interaction range. But on the other hand we know that GR has been tested in our
Solar System to very high accuracy which seems to exclude any
extra light degrees of freedom. The way out of this problem is to
have some mechanism that can decouple the extra degrees of freedom
in the Solar System. As explained in the previous section, for the
type of models (\ref{action}) the mass of the extra scalar grows
for large curvatures which suggests that such a mechanism could take
place. It is instructive to see how this happens and what
the modification will be for a spherically symmetric solution
corresponding to a central mass source $M$ on a cosmological background,
which we will take to be de Sitter for simplicity. (Notice that flat
Minkowski space typically won't be solution.) From
(\ref{runningmass}) we see that on the cosmic background the extra
scalar is indeed very light: $m_s\sim H_0 (\sim \mu)$. And at first
order in the weak field expansion the  Schwarzschild solution reads
\cite{Navarro:2005da} (for distances $r\ll H_0^{-1}$):  \bea ds^2
&\simeq& -\left(1-\frac{2G^{eff}_NM}{r}-
\frac{2G^{eff}_NM}{3r}\right)dt^2
+\left(1+\frac{2G^{eff}_NM}{r}-\frac{2G^{eff}_NM}{3r}\right)dr^2+r^2d\Omega^2_2
\label{ldsol}\,,\eea where we have written separately the
contributions from the spin 2 graviton and the scalar. If this was
the solution in the Solar system, the theory would be clearly ruled
out. Indeed, the contribution from the scalar in (\ref{ldsol}) would make
it impossible to fit both the orbits of the planets and the observed light
bending with the same value for Newton's constant. However, the weak
field expansion breaks down at a huge distance
$r_V=(G_NM/H_0^3)^{1/4}$. For the Sun this distance is of the order
of 10 kpc\footnote{ A more realistic set up for the Sun would be
to treat it as a probe on the background of the Milky Way, one then
finds $r_V\sim$10 pc.}. For shorter distances, inside the Solar
system for instance, one can not trust the perturbative solution
(\ref{ldsol}) anymore. This is in stark contrast with GR, where the
weak field expansion can be trusted throughout the whole Solar
System.

In fact we could have guessed the perturbative expansion to break
down when approaching the mass source, simply because we know that
we should recover something very close to the GR solution for short
distances. This happens because the curvature of the GR Schwarzschild solution
($Q=48(G_NM)^2/r^6$) blows up close to the mass source, thereby
killing the modification\footnote{We are assuming
a modification that contains the Kretschmann scalar $Q$. This
observation does not apply to $F(R)$
modifications. }. We can then estimate the
distance $r_c$ where the modification becomes important by looking
at the extra scalar. This scalar will produce an order one
modification for distances $r$, {\em smaller} than its inverse mass.
But we know that the mass depends on the background curvature (Eq.
\ref{runningmass}). For the GR Schwarzschild solution this gives a
mass that runs with the distance as: \be m_s(r)\sim \mu
\left(\frac{G_NM}{\mu^2 r^3}\right)^{n+1} \,. \ee So we can expect the scalar
to really modify things, only for distances for which: \be
r<\frac{1}{m_s(r)}\,\,\,\,\,\,\Rightarrow\,\,\,\,\,\,
r>r_c\equiv\left(\frac{(G_NM)^{n+1}}{\mu^{2n+1}}\right)^{\frac{1}{3n+2}}\,. \ee
For the Sun, $r_c$ is at least of the order of 10 pc (for $n\geq
1$), so the background dependence of the scalar mass indeed provides
a mechanism that decouples the scalar in the Solar System. We recover
therefore GR, up to small corrections. We have calculated these
corrections in an expansion on the GR solution
\cite{Navarro:2005gh}. They are typically smaller for larger values
of $n$ in (\ref{specmodels}) and their effects are too small for
detection.

So the situation for a static mass source on a cosmic background can
be summarized as follows. At ultra large distances we get the
perturbative solution (\ref{ldsol}) corresponding to an extra
massless scalar with, in addition, a rescaled Newton's constant
$G_N^{eff}$. For distances smaller than $r_V$ this perturbative
picture breaks down and we enter a non-perturbative regime. So far
we can not say that much about this regime, since the expansions that we
have used break down. But then, at  distances smaller than
$r_c(<r_V)$, the scalar decouples and we can expand on GR, finding a
solution that is very close to the GR one and where the corrections can be quantified.

Let us now put the centre of a galaxy as the mass source. One
could impose as a condition in these theories that they produce no modification in the
dynamics of the galaxy. This will be the case for large values of $n$ in
(\ref{specmodels}), that give $r_c \sim 1$Mpc. A more ambitious
alternative is to look for models that give a
modification that could simulate the effects of dark matter at the
galactic level. First of all, one can take a model for which
$G_N^{eff} > G_N$.  Measuring rotation curves at ultra large
distances, one would then infer the mass of the galaxy to be larger
than what it actually is. In that sense the non-perturbative region
indeed seems to have the characteristics of a dark matter halo.
However, from the success of MOND\cite{Sanders:2002pf}, we know
that the distance ($r_c$) where the would-be dark matter halo
begins should correspond to a universal acceleration $a_0\sim H_0$.
This is precisely the case for logarithmic actions\cite{Navarro:2005ux}. But due to lack of space we will have to
refer to the proceedings of another talk\cite{proceedings}, given by
one of us at a different session of this year's Rencontres de
Moriond, for a report on the interesting phenomenology of this class
of models.
\section{Conclusions}
Models of the type (\ref{action}) clearly have potential as a possible
alternative for $\Lambda$CDM. Obviously, for them to
become a real contender there are still a lot of questions that need
to be answered. What is for instance the effect of the chameleon-like behaviour of the gravitational excitations on the CMB? Another
concern, derived from the non-perturbative behaviour of these models,
is that it is not yet clear how to estimate the effects of
quantum corrections for this type of actions. This would be necessary, for instance,
to have an idea of the amount of fine tuning required in these
effective actions. Still, we believe that it is worthwhile to
explore the bottom-up approach of building a generally covariant classical action to
model long distance modifications of gravity; the advantage being
that one can make clear contact with experiment at many levels.
\section*{Acknowledgments}
We would like to thank the organizers of the Rencontres de Moriond
for inviting us to give a talk. K.V.A. is supported by a
postdoctoral grant of the Fund for Scientific Research Flanders
(Belgium). He would also like to thank the participants of the
cosmology session for stimulating discussions and interactions of a
more practical nature after an unpleasant encounter with one of the
slopes.

\end{document}